\documentclass[aps,twocolumn,floatfix]{revtex4-2}

\usepackage{amsmath,amsfonts,amssymb,braket,qcircuit,algorithm,algpseudocode,caption} 
\usepackage{graphicx}
\usepackage[]{units} 
\usepackage[T1]{fontenc}
\usepackage{hyperref} \hypersetup{ colorlinks=true, linkcolor=blue, filecolor=blue, urlcolor=blue, citecolor=blue }

\algnewcommand\algorithmicinput{\textbf{Input:}}
\algnewcommand\Input{\item[\algorithmicinput]}

\algnewcommand\algorithmicdefine{\textbf{Define:}}
\algnewcommand\Define{\item[\algorithmicdefine]}

\algnewcommand\algorithmicinitialise{\textbf{Initialise:}}
\algnewcommand\Initialise{\item[\algorithmicinitialise]}

\algnewcommand\algorithmicoutput{\textbf{Output:}}
\algnewcommand\Output{\item[\algorithmicoutput]}

\begin{document} 
\title{Quantum Computation by Spin Parity Measurements with Encoded Spin Qubits} 
\author{Matthew Brooks} 
\email{matthew.brooks@lps.umd.edu} 
\author{Charles Tahan} \affiliation{Laboratory for Physical Sciences, 8050 Greenmead Dr., College Park, MD 20740, USA}
 
\begin{abstract}
 
Joint measurements of two-Pauli observables are a powerful tool for both the control and protection of quantum information. By following a simple recipe for measurement choices, single- and two- qubit rotations using two-Pauli parity and single qubit measurements are guaranteed to be unitary whilst requiring only a single ancilla qubit. This language for measurement based quantum computing is shown to be directly applicable to encoded double quantum dot singlet-triplet spin qubits, by measuring spin-parity between dots from neighboring qubits. Along with exchange interaction, a complete, leakage free, measurement based gate set can be shown, up to a known Pauli correction. Both theoretically exact spin-parity measurements and experimentally demonstrated asymmetric spin-parity measurements are shown to be viable for achieving the proposed measurement based scheme, provided some extra leakage mitigating measurement steps. This new method of spin qubit control offers a leakage suppressed, low resource overhead implementation of a measurement-based control that is viable on current spin qubit processor devices.
	
\end{abstract}
\maketitle

\section{Introduction} \label{sec:Intro}

Measurement based quantum computing (MBQC) is a proposed method of processing quantum information that seeks to replace operations given by continuous control fields with discrete quantum measurements\cite{raussendorfOnewayQuantumComputer2002,raussendorfMeasurementbasedQuantumComputation2003,raussendorfFaulttolerantOnewayQuantum2006,browneOnewayQuantumComputation2006,raussendorf2007fault,MeasurementBasedControlled,karzigScalableDesignsQuasiparticlePoisoningProtected2016,brooks2021hybrid}. Generally, there are two routes with which this can be achieved. The first approach is by generating some large initial entangled state containing all the qubits of a given processor, i.e. a resource state, and performing a series of single qubit measurements around a given measurement axis, until the remaining unmeasured qubits of the processor are in the desired state\cite{raussendorfOnewayQuantumComputer2002,raussendorfMeasurementbasedQuantumComputation2003,raussendorfFaulttolerantOnewayQuantum2006,browneOnewayQuantumComputation2006,raussendorf2007fault,bellExperimentalDemonstrationGraph2014,brooks2021hybrid}. This method is particularly applicable to photonic processors\cite{bellExperimentalDemonstrationGraph2014} whereby entanglement generation is experimentally achievable, but two-qubit gates are challenging. The second method involves the use of joint measurements, a measurement about an axis of a shared property of two or more qubits, to achieve non-unitary entanglement by measurement\cite{MeasurementBasedControlled,karzigScalableDesignsQuasiparticlePoisoningProtected2016,beenakker2004charge,ionicioiu2007entangling,zilberberg2008controlled,freedman2021symmetry,brooks2021hybrid}. This method has been proposed as a method of braiding neighbouring topological Majorana qubits via measurement of charge parity between the qubits and a shared quantum dot\cite{MeasurementBasedControlled,karzigScalableDesignsQuasiparticlePoisoningProtected2016}. Additionally, it has also been shown to be applicable by measuring joint properties of solid state spin qubits\cite{beenakker2004charge,ionicioiu2007entangling,zilberberg2008controlled,brooks2021hybrid}. Both approaches can be shown to provide a universal set of single and two-qubit gates, at the cost of an at worse polynomial\cite{raussendorf2003measurement} number of ancilla qubits.

Recent work has demonstrated the power of multiple qubit joint-Pauli parity measurements in designing quantum error correcting (QEC) codes\cite{hastings2021dynamically,gidney2021fault}. Generally such measurements on and $N$-qubit processor can be written as

\begin{equation}
	\mathcal{P}(\nu,s)=\Gamma\left(\mathbb{I}^{\otimes N} + (-1)^s \bigotimes_{i=1}^N\sigma_{\nu_i}^i\right)
	\label{eq:P_general}
\end{equation}

\noindent where $\nu$ is  a vector of length $N$ with elements $\nu_i \in \{0,x,y,z\}$ that describe a particular joint measurement, $\sigma_\mu$ are the Pauli matrices with $\sigma_0=\mathbb{I}$, $\mathbb{I}$ is the identity matrix, $\Gamma$ is a normalising constant, and $s\in\{0,1\}$ is given by the either measurement of the positive ($s=0$) or negative ($s=1$) eigenstates of the desired observable. For example, in the case $N=2$ a joint-Pauli measurement of $\nu=\{z,z\}$ is equivalent to either the $\ket{\Psi^+}\bra{\Psi^+}=(\ket{00}\bra{00}+\ket{11}\bra{11})/2$ Bell state projection if $s=0$ or $\ket{\Phi^+}\bra{\Phi^+}=(\ket{01}\bra{01}+\ket{10}\bra{10})/2$ Bell state projection if $s=1$. Note that in this notation when $\nu_i=0$ this corresponds to performing no measurement on qubit $i$. A shorthand notation for (\ref{eq:P_general}) will be used whereby a measurement of the type $\nu=\{0,x,y,z\}$ is described as a $IXYZ$ parity check. Application of rounds of such measurements provide robust quantum memories without the need for fixed logical qubit space with only two-qubit joint-Pauli measurements\cite{hastings2021dynamically}, i.e. joint-measurements are only performed between two neighbouring qubits. Whilst some work on using such a measurement toolset for qubit control in Majorana\cite{MeasurementBasedControlled,karzigScalableDesignsQuasiparticlePoisoningProtected2016} and spin qubits\cite{beenakker2004charge,ionicioiu2007entangling} systems has been done, demonstration of how to use measurements of this type for arbitrary qubit control has not been fully characterised. Additionally, these previous studies tend to focus on just a single entangling measurement sequence, without consideration for universal control.

In this work we demonstrate that both single and two-qubit operations can be achieved by rounds of two-qubit joint-Pauli measurements and a single ancilla qubit. Then, by employing the physical observable of spin parity of spins in two neighboring QDs, we demonstrate how to achieve such control scheme in double quantum dot (DQD) spin qubits with only measurement and nearest-neighbour exchange interaction. Spin qubits offer fast and precise spin-spin operations via the exchange interaction\cite{mills2022two,xue2022quantum,noiri2022fast,weinstein2022universal} as well as long coherence times\cite{veldhorst2014addressable, veldhorst2015two}. They are however vulnerable to noise from control voltage fluctuations, known as charge noise\cite{yoneda2018quantum,connors2019low,chan2018assessment}. Charge noise may be neutralised somewhat by encoding of logical qubits in two or more spins across multiple dots\cite{russ2017three,burkard2021semiconductor,harvey2019spin,takeda2020resonantly,maune2012coherent,weinstein2022universal,blumoff2022fast}. Unfortunately, this also allows for a new form of error where the constituent spins of a qubit couple to states outside of the logical spin subspace, known as leakage\cite{burkard2021semiconductor,weinstein2022universal,wardrop2014exchange,klinovaja2012exchange}. The proposed joint-Pauli parity measurement control scheme offers leakage-suppressed operation of singlet-triplet encoded qubits at their charge noise resistant symmetric operating point (SOP)\cite{reed2016reduced}, without reliance on both external magnetic fields or magnetic field gradients for control.

In this paper, first in Sec.~\ref{sec:MBQC} a general condition on how to perform MBQC with joint-Pauli parity measurements is given. Then in Sec.~\ref{sec:SpinMBQC} the case of a MBQC with a physical parity observable is discussed, spin-parity measurements between encoded DQD spin qubits. This is explored with both an ideal case of an exact spin-parity measurement, and the lab case of an asymmetric spin-parity measurement. A proposed entangling gate is simulated in Sec.~\ref{sec:Fidelity}, investigating the effect of infidelity of exchange pulses and measurements. Finally the results are summarised in Sec.~\ref{sec:Discussion}.

\section{Joint-Pauli MBQC} \label{sec:MBQC}

Projective measurements of quantum states are non-unitary processes in nature. Therefore, when designing an MBQC scheme, a balance between the choice of measurements and ancill\ae~must be found to ensure that the overall imparted rotations on the data qubits are unitary. For consistency, we will treat the initialisation of the ancilla qubits as another measurement step, i.e. if an ancilla requires is initialised to the $\ket{0}$ state we will treat this as if that qubit has been measured in the Pauli-$z$ basis with a fixed outcome. For single and two-qubit unitaries using arbitrary joint-Pauli measurements, the requirements of measurement choices and number of ancilla are the same: only a single ancilla qubit is needed and each measurement step must be selected such that the chosen observables' multi-qubit Pauli operator does not commute with the observable of the previous measurement step. To clarify the requisite restrictions on the measurement choices, consider a two-qubit system. If the second qubit is an ancilla initialised to the $\ket{+}$ state, this is written as an $IX$ measurement. Then if a $ZX$ joint-measurement is performed between the two qubits, any data in the data qubit will be lost and the overall process cannot be made to be unitary since $[IX,ZX]=0$. However, if the joint-measurement of $ZZ$ is chosen instead, then the overall process can be made to be unitary as $[IX,ZZ]\neq0$.

This rule of measurement selection is shown here explicitly with single qubit unitaries. To perform a single qubit rotation with joint-Pauli measurements, the recipe is: two-qubit system, the first in some data state $\ket{\psi}$, the second qubit reserved as the ancilla. The ancilla is initialised, then a joint measurement on the two qubits creates an entangled pair, the ancilla is measured along a chosen single qubit basis to disentangle the pair such that some rotation $U$ is performed on the data qubit. Explicitly this sequence of measurements can be written as follows

\begin{equation}
	\begin{split}
			\Pi_{1-q}=\Gamma\left(\mathbb{I}^{\otimes 2}+\mathbb{I}\otimes\sigma_\zeta\right)\left(\mathbb{I}^{\otimes 2}+\sigma_\nu\otimes\sigma_\xi\right)&\\\left(\mathbb{I}^{\otimes 2}+\mathbb{I}\otimes\sigma_\mu\right)&
	\end{split}
	\label{P1q}
\end{equation}

\noindent assuming the outcomes of all measurements are of the positive eigenstate ($s=0$). So, reading (\ref{P1q}) from right to left the measurements are ancilla initialisation, entanglement, and ancilla disentangling with chosen non-identity Pauli observables $\mu$, $\nu-\xi$ and $\zeta$ respectively. To see the effect of the measurement sequence on the data qubit, we must trace over the ancilla qubit. Since the process should completely disentangle the ancilla qubit, (\ref{P1q}) can be written as 

\begin{equation}
	\Pi_{1-q}=\sum_{i=\{0,x,y,z\}} \gamma_i U\otimes\sigma_i
	\label{Psum}
\end{equation}

\noindent where $\gamma_i$ are constants. From expanding (\ref{P1q}) and assuming the form of (\ref{Psum}), $U$ is given as

\begin{equation}
	U=\Gamma\left[(1+\delta_{\zeta,\mu})\mathbb{I}+i^{\epsilon_{\zeta,\xi}}(\delta_{\zeta,\xi}+i^{\epsilon_{\zeta,\xi\oplus \mu}}\delta_{\zeta,\xi\oplus \mu})\sigma_\nu\right]
	\label{1QU}
\end{equation}

\noindent where $\delta_{i,j}$ is the Kronecker delta, $\varepsilon_{i,j}$ is the Levi-Civita symbol and the notation $i\oplus j$ is given from the product of the two Pauli matrices $\sigma_i \sigma_j=i^{\varepsilon_{i,j}}\sigma_{i\oplus j}$. Defining $1+\delta_{\zeta,\mu}=\alpha$ and $i^{\epsilon_{\zeta,\xi}}(\delta_{\zeta,\xi}+i^{\epsilon_{\zeta,\xi\oplus \mu}}\delta_{\zeta,\xi\oplus \mu})=\beta$, then the unitary condition for the rotation $U$ given by

\begin{equation}
	UU^\dagger=\mathbb{I}=|\Gamma|^2\left[(|\alpha|^2+|\beta|^2)\mathbb{I}+(\alpha\beta^*+\alpha^*\beta)\sigma_\nu\right]
\end{equation}

\noindent is only satisfied when 

\begin{equation}
	\alpha\beta^*=-\alpha^*\beta.
\end{equation}
\noindent Since $\alpha$ is always real this implies that $\beta=0$ or $\pm i$. These conditions are satisfied only when $\mu\neq \xi$ and $\xi \neq \zeta$, and therefore the joint-Pauli MBQC scheme only returns unitary rotations if and only if each measurement is selected such that it does not commute with the previous measurement chosen. From (\ref{1QU}) it is apparent that the scope of what single qubit unitaries that can be achieved in this scheme is limited to $\{I,S,S^\dagger,XH,HX,HSH,HS^\dagger H\}$, up to a local Pauli correction depending on the outcomes of each of the measurements. Unfortunately, while this does form a Clifford set, this does not make a universal set of single qubit gates, and so will need to be complemented by a $T$ gate, or some tunable non-Pauli observable such as

\begin{equation}
	\mathcal{P}_{W}(\theta,s)=\frac{1}{2}\left(\mathbb{I}+(-1)^s e^{I \theta \sigma_z /2}\sigma_x e^{-I \theta \sigma_z /2}\right)
\end{equation}

\noindent for completeness. Equally, this can also be achieved by initialising the ancilla in a state that is not an eigenstate of one of the Pauli matrices.

The same constraints on implementing single qubit gates apply to two-qubit gates. For two-qubit unitaries one needs only three physical qubits, the first a data qubit $\ket{\psi_1}$, the second an ancilla, and the third another data qubit $\ket{\psi_2}$. After the ancilla qubit is initialised, a two-qubit joint-measurement is done between one of the data qubits, say $\ket{\psi_1}$, and the ancilla, followed by joint-measurement between the other data qubit $\ket{\psi_2}$ and the ancilla, before finally the ancilla qubit is disentangled by a chosen measurement. Again, all measurements, including the initialisation, must obey the measurement selection rule outlined prior, such that the resulting gate on the data qubits is unitary. Two-qubit gates performed in this scheme can be both entangling or a product of two single qubit gates, depending on the choice of measurements, and will vary up to local Pauli corrections depending of the outcomes of each of the measurement steps. An example of an implementation of an entangling gate of this type has been previously discussed with respect to Majorana qubits\cite{karzigScalableDesignsQuasiparticlePoisoningProtected2016,MeasurementBasedControlled}. There, the sequence that was discussed, as limited by what is possible with the proposed architecture, $IZI\rightarrow ZXI\rightarrow IZX\rightarrow IXI$ which evidently obeys the measurement selection rule and is equivalent to a $CNOT$ gate on the data qubits\cite{MeasurementBasedControlled}.

\section{Spin-Parity MBQC} \label{sec:SpinMBQC}

Spin qubits are a natural fit for control by joint-Pauli measurement. Measurement of the parity of two spins, being either aligned or anti-aligned, is a $ZZ$ joint-measurement that has previously been discussed as a means of controlling single QD spin-1/2 qubits\cite{beenakker2004charge,ionicioiu2007entangling}. The spin-parity measurement can be written as follows

\begin{equation}
	\begin{split}
		\mathcal{P}_{\text{sp}}(s)=(1-s)\left(\ket{\uparrow\uparrow}\bra{\uparrow\uparrow}+\ket{\downarrow\downarrow}\bra{\downarrow\downarrow}\right)&\\+s\left(\ket{\uparrow\downarrow}\bra{\uparrow\downarrow}+\ket{\downarrow\uparrow}\bra{\downarrow\uparrow}\right)&
	\end{split}
	\label{PspinSym}
\end{equation}

\noindent where $s\in\{0,1\}$ such that $s=0$ measures the spins as aligned, while $s=1$ measures the spins as anti-aligned. Assuming access to single spin Hadamard gates an entangling $CZ$ gate can be implemented by a $IZI\rightarrow ZXI\rightarrow IZZ\rightarrow IXI$ measurement sequence\cite{ionicioiu2007entangling}. This sequence, similar to the Majorana sequence, follows the outlined measurement selection rule. However, this scheme abstracts the spin parity measurement such that in a physical implementation of the scheme, two non measurement-based $CNOT$ gates and an additional qubit is needed for each parity measurement. Additionally, as this sequence requires single spin-1/2 qubit gates by ESR-driven pulses, the system is susceptible to charge noise.

\subsection{Exact Spin-Parity Measurements} \label{sec:Symmetric}

Here instead we will consider DQD single-triplet qubits. The logical qubit states are a subset of the full spin space of a DQD whereby the singlet $\ket{S}=(\ket{\uparrow\downarrow}-\ket{\downarrow\uparrow})/\sqrt{2}=\ket{0}$ and the triplet-0 $\ket{T_0}=(\ket{\uparrow\downarrow}+\ket{\downarrow\uparrow})/\sqrt{2}=\ket{1}$. These encoded qubits can be tuned to be resistant to first-order charge noise as well as offer accurate and fast single qubit phase gates by varying the exchange interaction between the two spins. However, universal control of such qubits requires additional on chip engineering such as magnetic field gradients by either variable g-factors or micro-magnets. Previous work on parity measurement based gates in such encoded spin qubits have relied on gate based full single qubit control\cite{zilberberg2008controlled}, specifically $H$ gates, which demand such considerations in chip design. These requirements can potentially limit future scalability.

Suppose instead there are two of such qubits in a linear array, consisting of four identical neighouring QDs. The first two dots encoded some state in the singlet-triplet subspace $\ket{\psi_1}=\alpha_1 \ket{S}+\beta_1 \ket{T_0}$ and the last two encode some state $\ket{\psi_2}=\alpha_2 \ket{S}+\beta_2 \ket{T_0}$. If a spin-parity measurement is done between the second and third dot, i.e. one dot from each constituent DQD qubit, the effect of the measurement in the qubit space, is an $XX$ joint-Pauli measurement. Assuming there was no leakage in the preparation of the two qubit states, the spin-parity measurement will not couple to states outside singlet-triplet subspace.

\begin{figure}
	[t]
	 \[\hspace{-5em}\Qcircuit @C=0.8em @R=1em @!R {
 	 \lstick{\ket{\psi}} & \qw & \gate{\frac{\pi}{2}} & \multigate{1}{\mathcal{P}_{\text{sp}}(s_1)} & \qw &  \gate{\frac{-(-1)^{s_1+s_2}\pi}{2}} & \qw &\rstick{X^{s_1+s_2}H\ket{\psi}} \\ \lstick{\ket{+}} & \qw & \gate{\frac{\pi}{2}} & \ghost{\mathcal{P}_{\text{sp}}(s_1)} & \qw & \measureD{\mathcal{M}_{ST} (s2)} }\]
	\caption{Circuit diagram of the exchange gates and measurement sequence needed to achieve a single qubit Hadamard up to a local Pauli-$x$ correction. All single qubit gates here are exchange phase gates of angles indicated. 
	} 
	\label{fig:single_qubit_sym}
\end{figure}
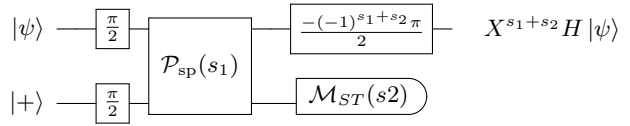

 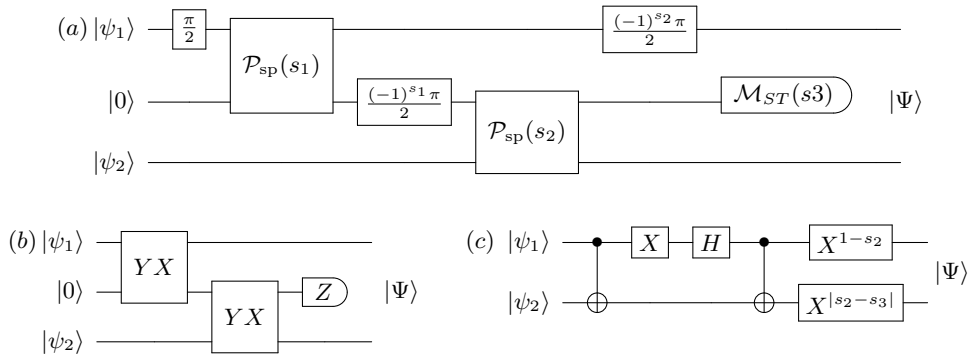
\begin{figure*}[t]
 			 \[\Qcircuit @C=1em @R=1em {
 		 	\lstick{(a)\ket{\psi_1}} & \gate{\frac{\pi}{2}} & \multigate{1}{\mathcal{P}_{\text{sp}}(s_1)} &\qw &\qw  & \gate{\frac{(-1)^{s_2}\pi}{2}}  & \qw &\qw  &\qw  
 		    \\ \lstick{\ket{0}} & \qw &\ghost{\mathcal{P}_{\text{sp}}(s_1)} &\gate{\frac{(-1)^{s_1}\pi}{2}} & \multigate{1}{\mathcal{P}_{\text{sp}}(s_2)} & \qw & \measureD{\mathcal{M}_{ST} (s3)}& \rstick{\ket{\Psi}} 
 		\\ \lstick{\ket{\psi_2}} & \qw & \qw &\qw & \ghost{\mathcal{P}_{\text{sp}}(s_2)} &\qw &\qw &\qw &\qw \\ \\ 
 }\]
 \begin{tabular}{c@{\hskip 1in} c}
 				 \Qcircuit @C=1em @R=1em {
 			 	\lstick{(b)\ket{\psi_1}}  & \multigate{1}{YX} &\qw &\qw  &\qw  
 			    \\ \lstick{\ket{0}} &\ghost{YX}  & \multigate{1}{YX}  & \measureD{Z}& \rstick{\ket{\Psi}} 
 			\\ \lstick{\ket{\psi_2}} &\qw & \ghost{YX} &\qw  &\qw  
 	} & 
 	 \Qcircuit @C=1em @R=1em {
  	\lstick{(c)\text{        }\ket{\psi_1}} &\ctrl{1} & \gate{X}& \gate{H}  & \ctrl{1}  & \gate{X^{1-s_2}} & \qw &\raisebox{-3em}{$\ket{\Psi}$} 
     \\ \lstick{\ket{\psi_2}} & \targ &\qw&\qw & \targ  & \gate{X^{|s_2-s_3|}} & \qw}
 \end{tabular}
 	\caption{(a) Circuit diagram of the phase gates and spin-parity measurement sequence needed to achieve a two qubit entangling unitary. (b) Equivalent circuit diagram as in (a) written with joint-Pauli parity measurements. (c) Standard Clifford gate circuit diagram of the two qubit unitary performed by the measurement and phase sequence given in (a) with Pauli-$x$ corrections.}
 	\label{fig:two_qubit_sym}
 \end{figure*}

Along with the native exchange based phase gates, the $XX$ parity measurement from the spin-parity measurement of neighboring dots is sufficient to design a complete Clifford+T gate set or single and entangling gates. Application of phase $\pm \pi/2$-gates before and after the joint measurement gives access to the following set of joint measurements $\{XX,XY,YX,YY\}$. Therefore, measurement sequences that give unitary rotations may be implemented by following the previously outline measurement selection rule. The remaining ingredients needed are choice of initialisiation of ancilla qubits and the final disentangling single qubit measurements. Initialisation of ancill\ae~is flexible, as both $z$-initialisation $\ket{0}=\ket{S}$ and $x$-initialisation $\ket{+}=(\ket{S}+\ket{T_0})\sqrt{2}=\ket{\uparrow\downarrow}$ are achieved by either non-adiabatic tunneling from a $(2,0)$ to a $(1,1)$ charge regime, or adiabatic tunneling respectively\cite{philips2022universal}. Consequently, with exchange based phase gates $y$-initialisation $\ket{+i}=(\ket{S}+i \ket{T_0})\sqrt{2}=(\ket{\uparrow\downarrow}+i \ket{\downarrow\uparrow})\sqrt{2}$ is also achievable. Equally, this technique can be used to inialise an ancilla state for a non Clifford gate by selecting a phase that is not an eigenstate of either Pauli-$x$ or $y$, but rather somewhere inbetween. The choice of final disentangling measurements is limited by experiment to measurement along the Pauli-$z$ axis, i.e. a measurement of the singlet-triplet spin state of the qubit

\begin{equation}
	\mathcal{M}_{ST}(s)=|s-1| \ket{S}\bra{S} + s \sum_{i=0,\pm}\ket{T_i}\bra{T_i}
\end{equation}

\noindent where measurement outcome $s=0$ projects on the singlet state, and outcome $s=1$ projects onto the triplet spin space. This can be achieved by either conventional Pauli spin blockade (PSB)\cite{petta2005coherent,burkard2021semiconductor} or by coupling to a superconducting resonator\cite{ruskov2019quantum}. As the entangling spin-parity measurements prior to the disentangling measurement do not couple to leaked $\ket{T_\pm}=\ket{\uparrow\uparrow}(\ket{\downarrow\downarrow})$ spin states, measurement of the triplet spin space only projects onto $\ket{T_0}$ state making the disentanglement step an exact Pauli-$z$ measurement.

An example of how to perform a Hadamard single qubit gate by spin-parity measurement and exchange alone is given in Fig.~\ref{fig:single_qubit_sym}. This sequence is effectively an $IY\rightarrow YX\rightarrow IZ$ measurement sequence. Assuming a spin-parity measurement of the form (\ref{PspinSym}), at no point in this sequence do either of the two qubits couple to leaked spin states outside the qubit subspace, and so this scheme can be considered leakage-free. Note however that after the sequence, there is a measurement outcome dependent Pauli-$x$ correction. This is a common consequence of MBQC and can either be remedied by assuming non-measurement based single qubit gates, which in this case will require the re-introduction of a magnetic field gradient, or by considered post-processing of the final state tomography data. An example of one such possible entangling gate with the proposed scheme is given in Fig.~\ref{fig:two_qubit_sym} (a) with the corresponding equivalent gate based circuit given in Fig.~\ref{fig:two_qubit_sym} (c). This measurement sequence is effectively equivalent to an $IZI\rightarrow YXI\rightarrow IYX\rightarrow IZI$ sequence, as shown in Fig.~\ref{fig:two_qubit_sym} (b), optimised however to require a minimum number of phase gates. Although the resulting gate is an unusual entangling primitive to operate a quantum computer with, it is a maximally entangling gate, and generating Bell-pairs from classical inputs in the following sequence $\ket{00}\rightarrow\ket{\Psi^-}$, $\ket{01}\rightarrow\ket{\Phi^-}$, $\ket{10}\rightarrow\ket{\Phi^+}$, and $\ket{11}\rightarrow\ket{\Psi^+}$. Note that this is not the only example of an entangling two qubit unitary that can be achieved with this architecture, as other rotations can be achieved by selecting different patterns of phase gates.

\subsection{Asymmetric Spin-Parity Measurements} \label{sec:Asymmetric}

Entangling parity measurement based gates in encoded spin qubits similar to those given so far have previously been discussed assuming exact spin-parity measurements and single qubit Hadamard gates\cite{zilberberg2008controlled}. Experimentally, however, spin-parity measurements are achieved by tuning the energy levels of the two neighboring QDs such that the $\ket{S}$ and $\ket{T_0}$ states quickly decays to the (2,0)/(0,2) charge configuration singlet state $\ket{S_{L/R}}$ in the left/right dot. This therefore only leaves the spin-aligned triplet states $\ket{T_\pm}$ in the (1,1) charge configuration\cite{seedhouse2020parity,philips2022universal}. While this does measure if the two spins are aligned or anti-aligned, this method lacks the symmetry of an exact spin-parity measurement. This asymmetric spin-parity measurement may be written as

\begin{equation}
	\mathcal{P}_{ST}(s)=|s-1|(\ket{T_+}\bra{T_+}+\ket{T_-}\bra{T_-})+s \ket{S}\bra{S}.
	\label{eq:AsymPMeas}
\end{equation}

\noindent It is however worth noting that, although decay from the $\ket{T_0}$ to $\ket{S_{L/R}}$ is inherent, without some finite Zeeman energy splitting between the $\ket{T_0}$ and $\ket{S}$ the decay is relatively slow, and its rate uncontrollable. By re-introducing such a splitting, the rate of decay has been shown to be increased by four orders of magnitude\cite{seedhouse2020parity}.

	\begin{figure*}[t]
				 \[\Qcircuit @C=1em @R=1em {
			 	\lstick{(a)\ket{\psi_1}} & \gate{\frac{\pi}{2}} & \multigate{1}{\mathcal{P}_{ST}(s_1)} & \multigate{1}{R_{\mathcal{P}_{ST}}(s_1)} &\qw & \qw & \gate{\frac{(-1)^{s_1+s_2}\pi}{2}}  &\qw  &\qw &\qw  
			    \\ \lstick{\ket{0}} & \qw &\ghost{\mathcal{P}_{ST}(s_1)}  & \ghost{R_{\mathcal{P}_{ST}}(s_1)} &\gate{\frac{(-1)^{s_1}\pi}{2}} & \multigate{1}{\mathcal{P}_{ST}(s_2)}  & \qw & \measureD{\mathcal{P}_{ST} (s_3)}& \multigate{1}{R_{ST}(s_3)} & \rstick{\ket{\Psi}}
			\\ \lstick{\ket{\psi_2}} & \qw & \qw & \qw  &\qw & \ghost{\mathcal{P}_{ST}(s_2)} & \qw & \qw &\ghost{R_{ST}(s_3)} &\qw  
	}\]
	
	\[\Qcircuit @C=1em @R=1em { \\ \\
		 	\lstick{(b)\text{   }\ket{\text{Data}}}  &\multigate{1}{R_{\mathcal{P}_{ST}}} & \qw & \raisebox{-2em}{=} & &  \gate{\pi} & \multigate{1}{\mathcal{P}_{ST}(1)} & \qw & \raisebox{-2em}{$\cdots$} & & \multigate{1}{\mathcal{P}_{ST}(0)} & \qw\\
			  \lstick{\ket{\text{Ancilla}}} &\ghost{R_{\mathcal{P}_{ST}}} & \qw & & & \qw & \ghost{\mathcal{P}_{ST}(1)} & \qw & & & \ghost{\mathcal{P}_{ST}(0)} & \qw   }\]
			   
	  		 \[\Qcircuit @C=1em @R=1em {\\ \\
	  		 	 &\multigate{4}{R_{ST}} & \qw & \raisebox{-7em}{=} & & \qw & \multimeasureD{1}{S} &  \\
	  			 \lstick{\raisebox{2em}{$(c)\ket{\text{Ancilla}}$}}&\ghost{R_{ST}} & \qw & & & \qswap & \ghost{S} &  \\  & &  & & &  \qwx & &  &  \\
				   &\ghost{R_{ST}} & \qw & & & \qswap \qwx & \qw & \qw & \qw \\
				   \lstick{\raisebox{2em}{$\ket{\text{Data}}$}} &\ghost{R_{ST}} & \qw & & & \qw & \qw  & \qw & \qw \\ }\]
		\caption{(a) Circuit diagram of the phase gates, asymmetric spin-parity measurements and repeat measurement sequence needed to achieve an equivalent two qubit entangling unitary as in Fig.~\ref{fig:two_qubit_sym}(c). (b) Circuit diagram of the repeat entangling measurement sequence. (b) Circuit diagram of the repeat disentangling measurement sequence. Note that in (a) and (b) each wire denotes a qubit, comprised of 2 QDs, while in (c) each wire denotes a single QD, as to show the exchange $\pi$-pulse (swap gate) needed to enable repeat measurements.}
		\label{fig:two_qubit_Asym}
	\end{figure*}
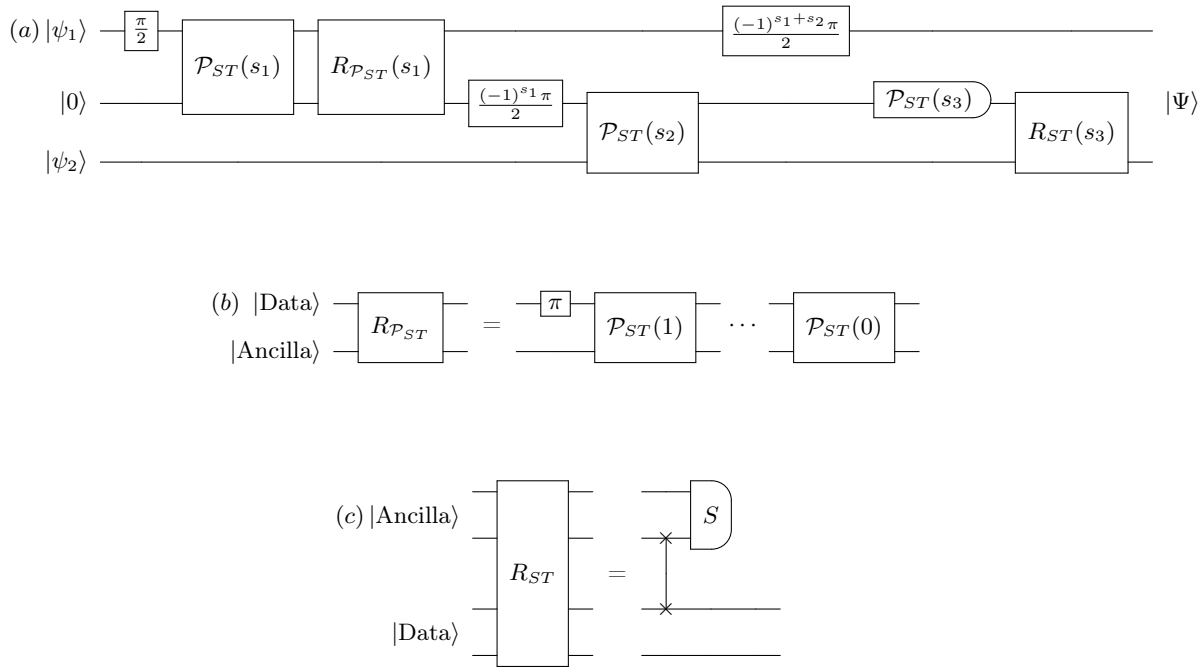

Assuming access to a measurement of this form, some considerations must be made. Firstly, let it be assumed that for ease of experimental implementation and consequently process efficiency, all measurement types used in this protocol are identical, i.e. they all asymmetric spin-parity measurements as in Eq.~\ref{eq:AsymPMeas}, including the final disentangling measurement. Secondly, when entangling two neighboring qubits by measurement, leakage may occur if a spin anti-aligned ($\ket{S}$) state is detected. To address this issue, some additional ingredients are needed to ensure that the final rotation of a given measurement sequence is unitary. If an entangling measurement detects a singlet state, causing leakage, this can be remedied by employing repeat measurements. By applying a $\pi$-phase gate to a select qubit, the same qubits can then be measured again until they are found to be spin-aligned. This, naturally, increases the measurement and exchange gate overhead of the overall rotation in an uncontrolled and probabilistic manner. However, such repeated measurements are not always needed. In the case of a two qubit gate, as in the Fig.~\ref{fig:two_qubit_Asym}, repeat measurements are required if, the first entangling measurement projects onto a singlet state, or if the disentangling measurement of the ancilla projects onto triplet space. In the first case, when the first entangling measurement projects onto the leaked state, the measurement can be repeated after application of a $\pi$-phase gate to the $\ket{\psi_1}$ data qubit until a triplet state is observed. Measurement of the triplet after a repeat measure recovers the desired unitary, up to a local $\pi$-phase (Pauli-$z$) correction on the $\ket{\psi_1}$ data qubit, which is accounted for later in the protocol. This type of repeat measurement is show in Fig.~\ref{fig:two_qubit_Asym}(b). The second case, when the disentangling measurement projects onto a triplet state, requires a repetition of the disentangling measurement on the ancilla qubit to regain the desired unitary. Here, however, a $\pi$-phase gate between one of ancilla qubits QDs and one of the $\ket{\psi_2}$ data qubit QDs is employed. This is a gate acting outside of the qubit subspace, and is equivalent to a SWAP-gate between the QDs in a spin-1/2 qubit picture.  This type of repeat measurement is shown in Fig.~\ref{fig:two_qubit_Asym}(c). Note that if both entangling measurements project onto the spin aligned triplet states, even after repeat measurements, then no leakage states are populated at the point of the disentangling step and the probability of measuring the spin aligned outcome is $0$. The necessity repeat of measurements is given in Tab.~\ref{tab:RepeatTab}. 

Although the outcome of quantum measurements are probabilistic, they can be weighted for efficiency. Due to the choice of the ancilla qubits initial state in the gate outline in Fig.~\ref{fig:two_qubit_Asym}, the probability of the first parity measurement projecting onto the leakage inducing spin anti-aligned ($\ket{S}$) state, is $1/3$, regardless of the input state $\ket{\psi_1}$. Therefore the requirement for repeating this measurement converges as $(1/3)^n$ where $n$ is the number of times the measurement has already been repeated. The disentangling measurement is also weighted regardless of input states, in such a way that only a maximum of one repetition is ever needed. This is due to the choice of disentangling by the asymmetric spin-parity measurement as opposed to a singlet-triplet qubit subspace measurement. Assuming the outcomes of the previous measurements allow for some leakage, and regardless of input state the probability of initially projecting to the spin aligned ($\ket{T_\pm}$) state is $2/3$. Whilst this is weighted such that a repeat measurement is likely, after a single implementation of the repeat measurement scheme, the probability of projecting to the spin anti-aligned ($\ket{S}$) state, and thus completing the two qubit unitary, is always $1$. Algorithm~\ref{alg:Repeats} details how both of the repeat measurement cycles are to be used.

\begin{algorithm}[H]
\caption{Two-qubit gate with Repeat Measurements}
\begin{algorithmic}

\Define $(D_1, D_2)\rightarrow Q_1$, $(D_3, D_4)\rightarrow A$, $(D_5, D_6)\rightarrow Q_2$
\Input Data qubits $Q_1\rightarrow\ket{\psi_1}$ and $Q_2\rightarrow\ket{\psi_2}$
\Initialise Ancilla $A$ $\rightarrow\ket{0}$

\State Exchange Pulse: $\frac{\pi}{2} \rightarrow Q_1$

\State Parity Measurement: $Q_1 \leftrightarrow A$ ($D_2,D_3$), output $s_1$

\If{$s_1=1$}
\State $s_{\text{Repeats}}=1$
  \While{$s_{\text{Repeats}}>0$}
    \State Exchange Pulse: $\pi \rightarrow Q_1$
    \State Parity Measurement: $A \leftrightarrow Q_1$ ($D_1,D_2$), update output $s_{\text{Repeats}}$
  \EndWhile
\EndIf

\State Exchange Pulse: $\frac{(-1)^{s_1}\pi}{2} \rightarrow A$

\State Parity Measurement: $A \leftrightarrow Q_2$ ($D_4,D_5$), output $s_2$

\State Exchange Pulse: $\frac{(-1)^{s_1+s_2}\pi}{2} \rightarrow Q_1$

\State Disentangling Parity Measurement: $A$ ($D_3,D_4$), output $s_3$

\If{$s_3=0$}
    \State Exchange Pulse: $\pi \rightarrow (D_4, D_5)$
    \State Disentangling Parity Measurement: $A$ ($D_3,D_4$)
\EndIf

\Output $Q_1 \otimes Q_2 =\ket{\Psi}$

\end{algorithmic}
\label{alg:Repeats}
\end{algorithm}

Some further improvement to the gate protocol in Fig.~\ref{fig:two_qubit_Asym} may be achieved by replacing the second entangling parity measurement and final disentangling parity measurements with initialising to the singlet state. This can be done by tuning the dots of that given measurement such that within a reasonable time ($\mathcal{O}(\unit[1]{\mu s})$) the two electrons will decay into a singlet state with $(0,2)/(2,0)$ charge configuration. Locking these measurement outcomes by this re-initialisation step improves the gate in two ways. Firstly, this eliminates the need for Pauli corrections to the final qubit state after the protocol as the gate in Fig.~\ref{fig:two_qubit_sym}(c) is locked such that $s_2=s_3=1$. Secondly, the need to ever repeat the final disentangling measurement is eliminated as decay to the singlet state will correct for any leakage accumulated. This will on average decrease the average exchange gate and measurement overhead and therefore the average gate time.

In recent work\cite{philips2022universal}, this asymmetric spin-parity measurement technique has been combined with adiabatic re-initialisation of the (1,1) from the (2,0)/(0,2) charge configuration $\ket{S_{L/R}}\rightarrow\ket{\uparrow\downarrow}$. While this allows for perfect spin-parity measurements within the qubit space of a DQD encoded qubit, when employing an asymmetric measurement to entangle two of such qubits together, the adiabatic initialisation step can effectively be treated as a second measurement in spin space that leads to loss of encoded information. This is a direct consequence of the violation of the commutation rule outlined prior. As such, the protocols described require that the $\ket{S}$ in the (1,1) charge configuration is initailased after a $\ket{S_{L/R}}$ state is measured. 

\begin{table}
	[t]
	\begin{tabular} {| l || c | c |}
		\hline
		Measurement Outcome ($s_1,s_2,s_3$) & $R_{\mathcal{P}_{ST}}$ & $R_{ST}$\\
		\hline \hline
		(0,0,0) & No & - \\
		\hline
		(0,0,1) & No & No \\
		\hline
		(0,1,0) & No & No \\
		\hline
		(0,1,1) & No & \bf{Yes} \\
		\hline
		(1,0,0) & \bf{Yes} & - \\
		\hline
		(1,0,1) & \bf{Yes} & \bf{Yes} \\
		\hline
		(1,1,0) & \bf{Yes} & No \\
		\hline
		(1,1,1) & \bf{Yes} & \bf{Yes} \\
		\hline
	\end{tabular}
	\caption{Table of the 8 possible measurement outcomes ($s_1,s_2,s_3$) of the MBQC sequence proposed in Fig.~\ref{fig:two_qubit_Asym}(a) with the corresponding required repeat measurement types, $R_{\mathcal{P}_{ST}}$ the repeat entangling measurement and $R_{ST}$ the repeat disentangling measurement, to achieve a leakage free entangling two-qubit unitary. Note that entries of `-' have measurement probability of $0$, and are therefore not considered.}
	\label{tab:RepeatTab}
\end{table}

\section{Entangling Gate Fidelity and Leakage}

\label{sec:Fidelity}

\begin{figure}
	[t]
	\includegraphics[width=0.48\linewidth]{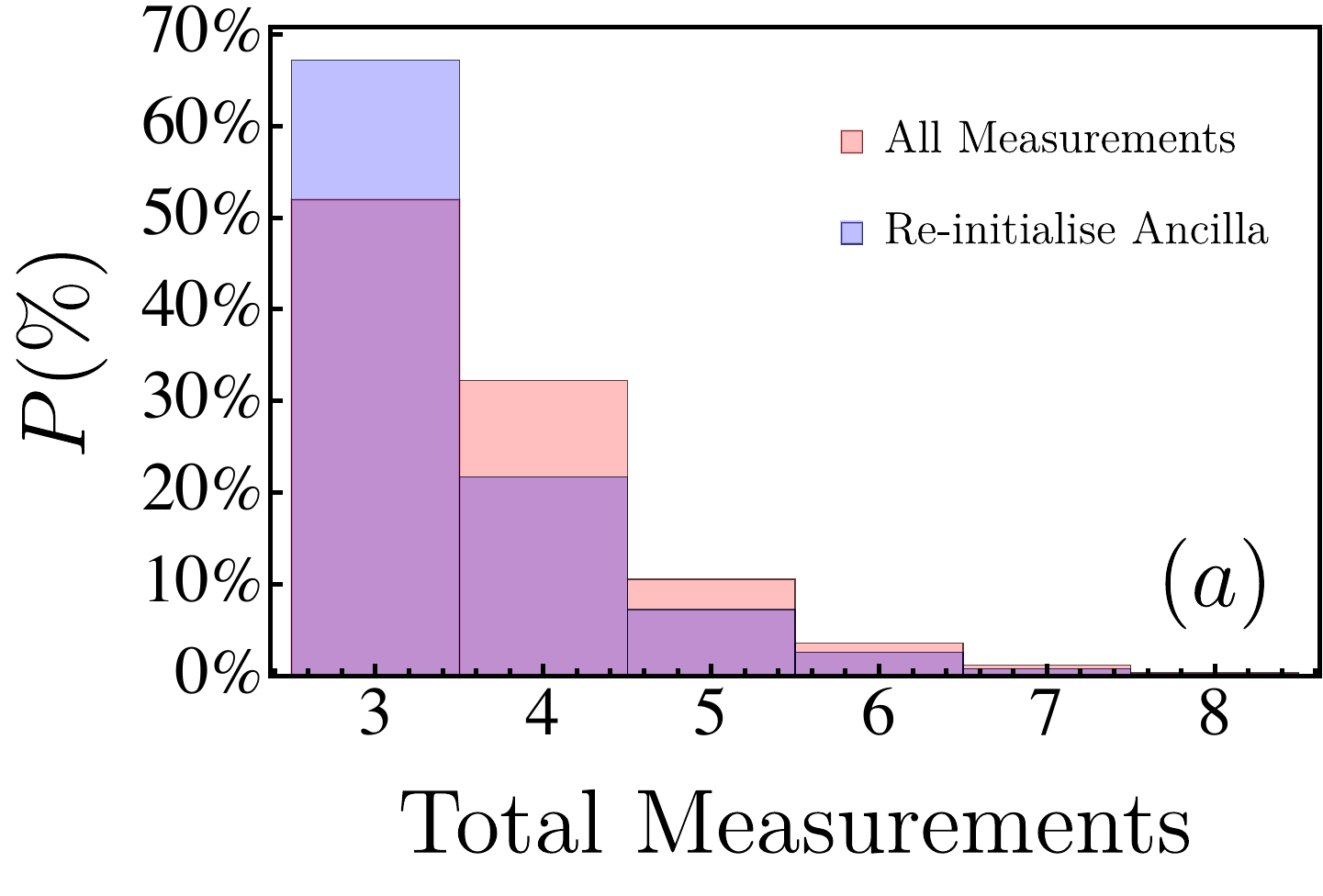}
	\includegraphics[width=0.48\linewidth]{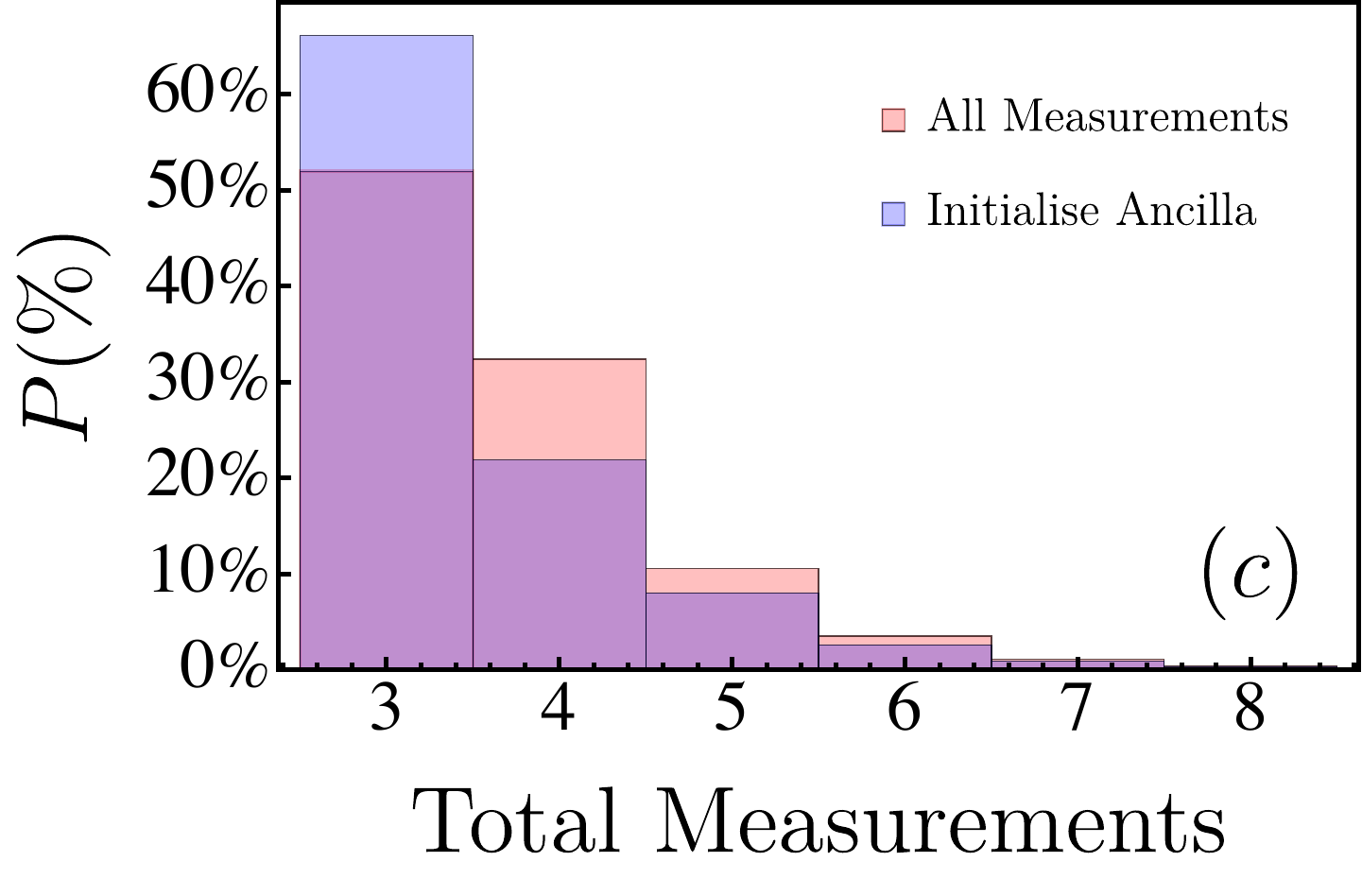}
	\includegraphics[width=0.48\linewidth]{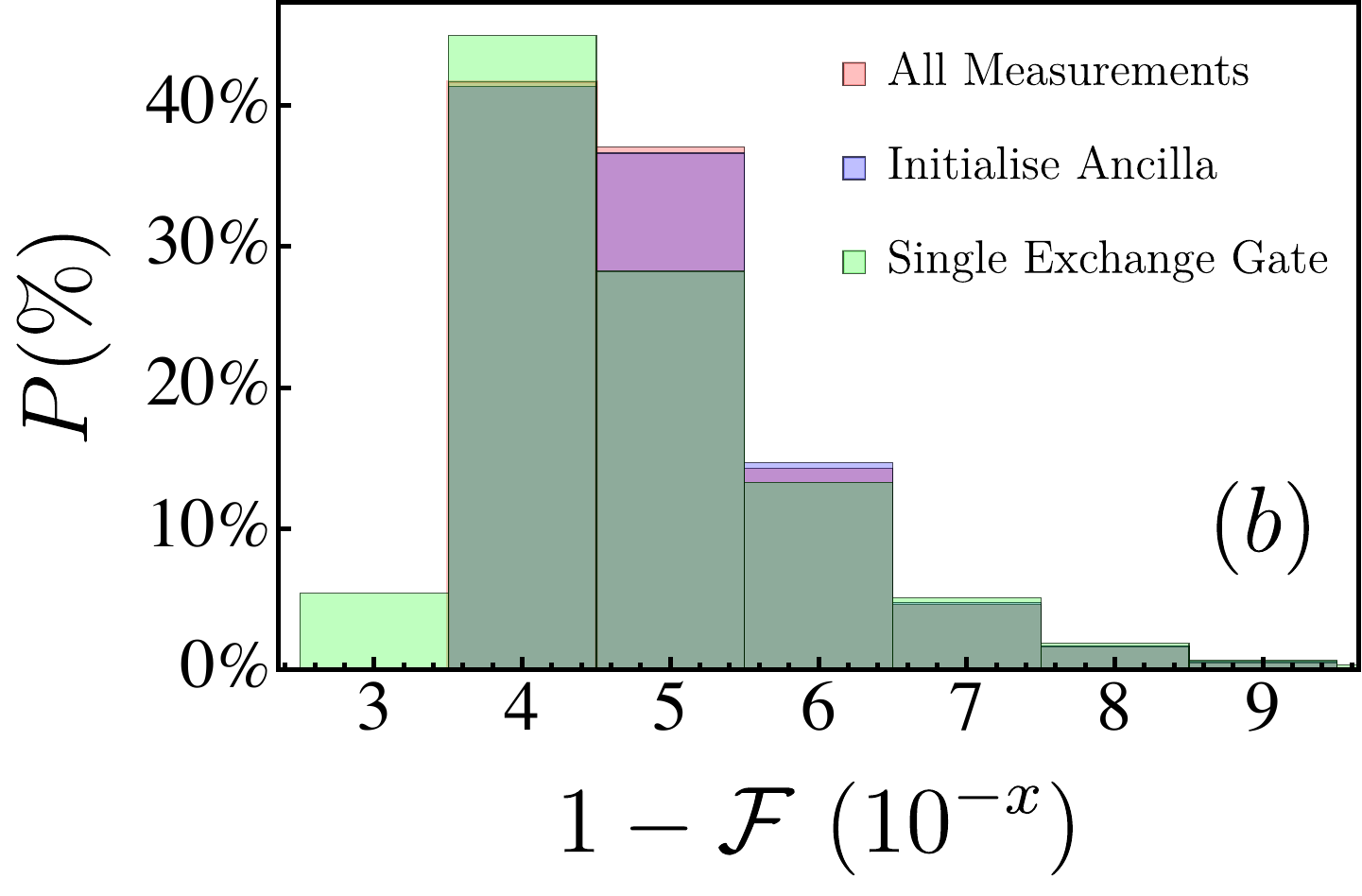}
	\includegraphics[width=0.48\linewidth]{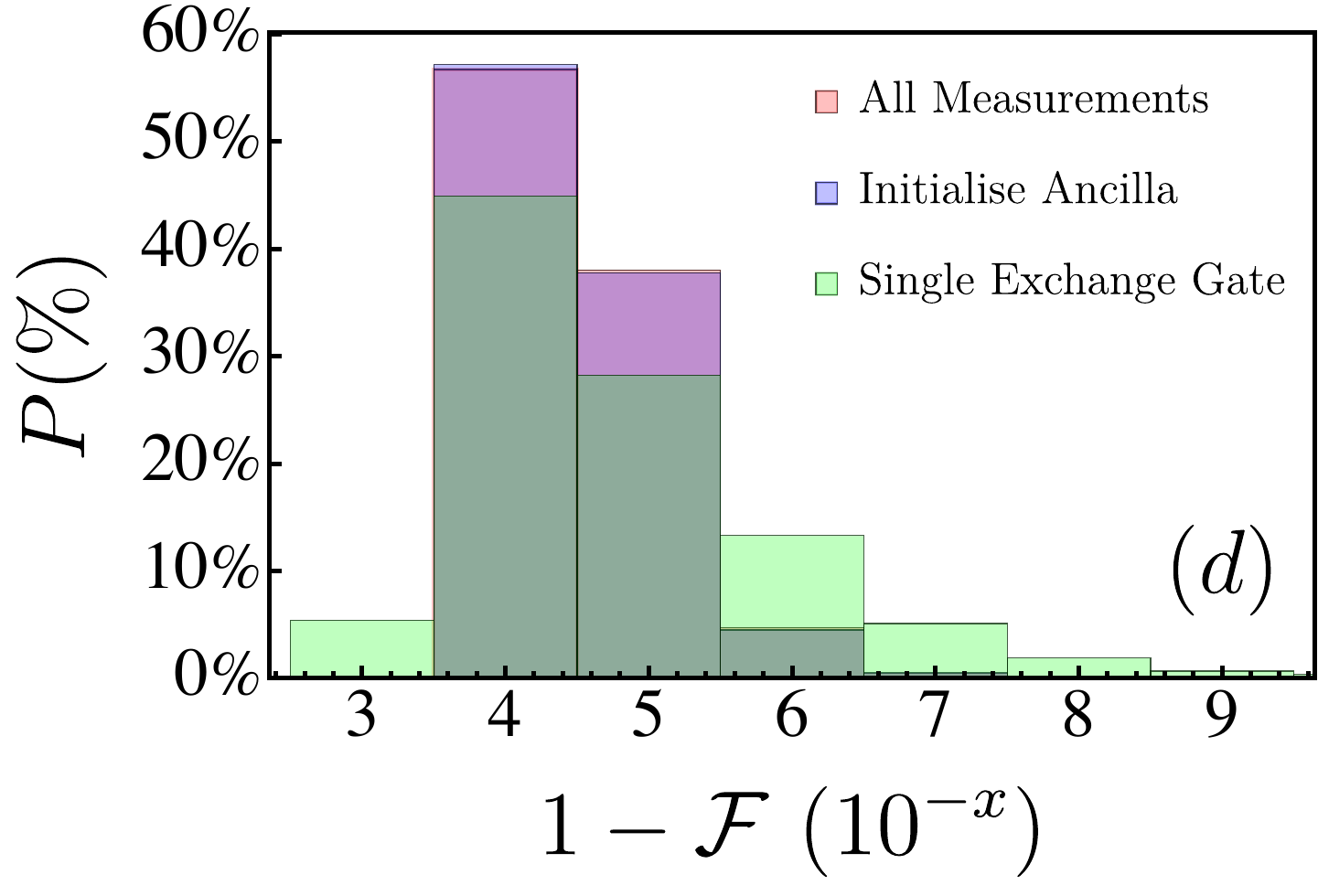}
	\caption{Histogram data of simulations of the proposed two qubit parity measurement based entangling gate with exchange pulse errors. Each data set is derived from $10^4$ iterations of the simulation and a random $1\%$ error in the exchange pulses ($\varepsilon^{\text{max}}_J=0.01$). (a) Percentage distribution of number of measurements needed to complete the entangling gate for the input state $\ket{\psi_1 \psi_2}=\ket{SS}$ for when all measurement steps are employed as measurements (red), and when the final measurement step is replaced by the discussed initialisation by decay (blue). (b) The infidelity distribution of the entangling gate for the input state $\ket{\psi_1 \psi_2}=\ket{SS}$ compared to the infidelity distribution of a single exchange pulse with the same simulated errors. (c) Percentage distribution of number of measurements needed to complete the entangling gate for random the input states. (d) The infidelity distribution of the entangling gate for random the input states} 
	\label{fig:SimulationHistograms}
\end{figure}
 
\begin{figure}
	[b]
	\includegraphics[width=0.48\linewidth]{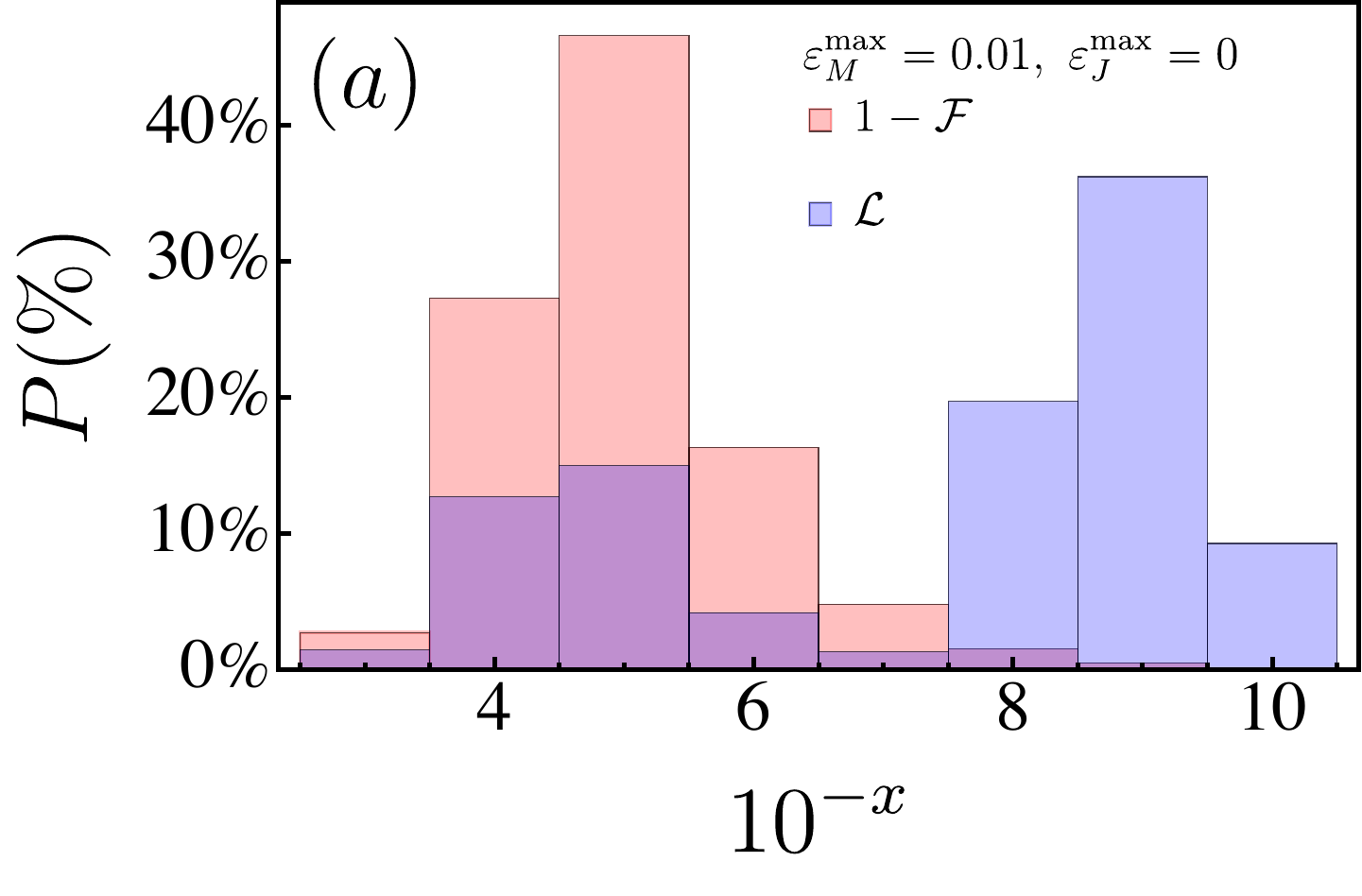}
	\includegraphics[width=0.48\linewidth]{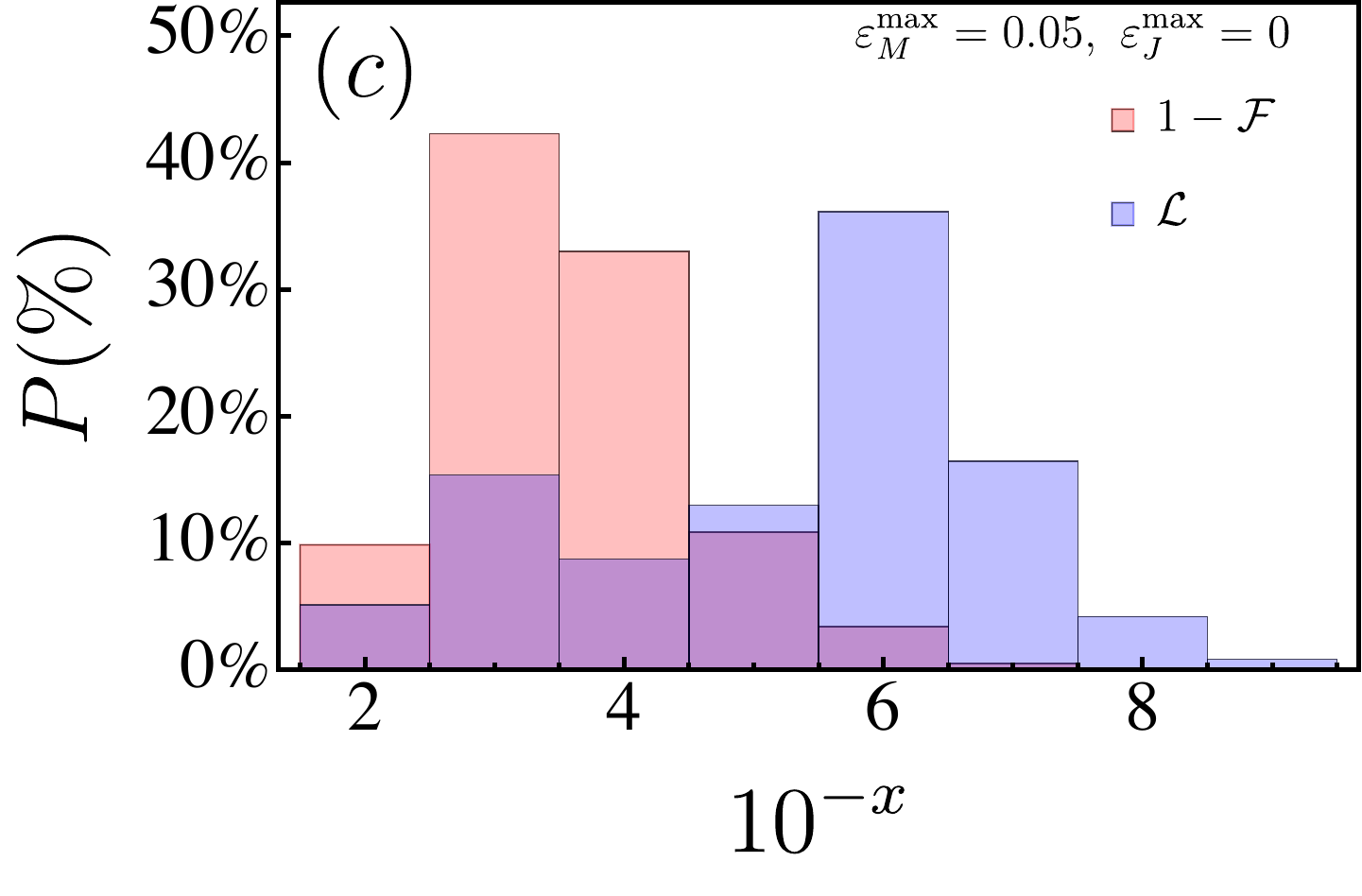}
	\includegraphics[width=0.48\linewidth]{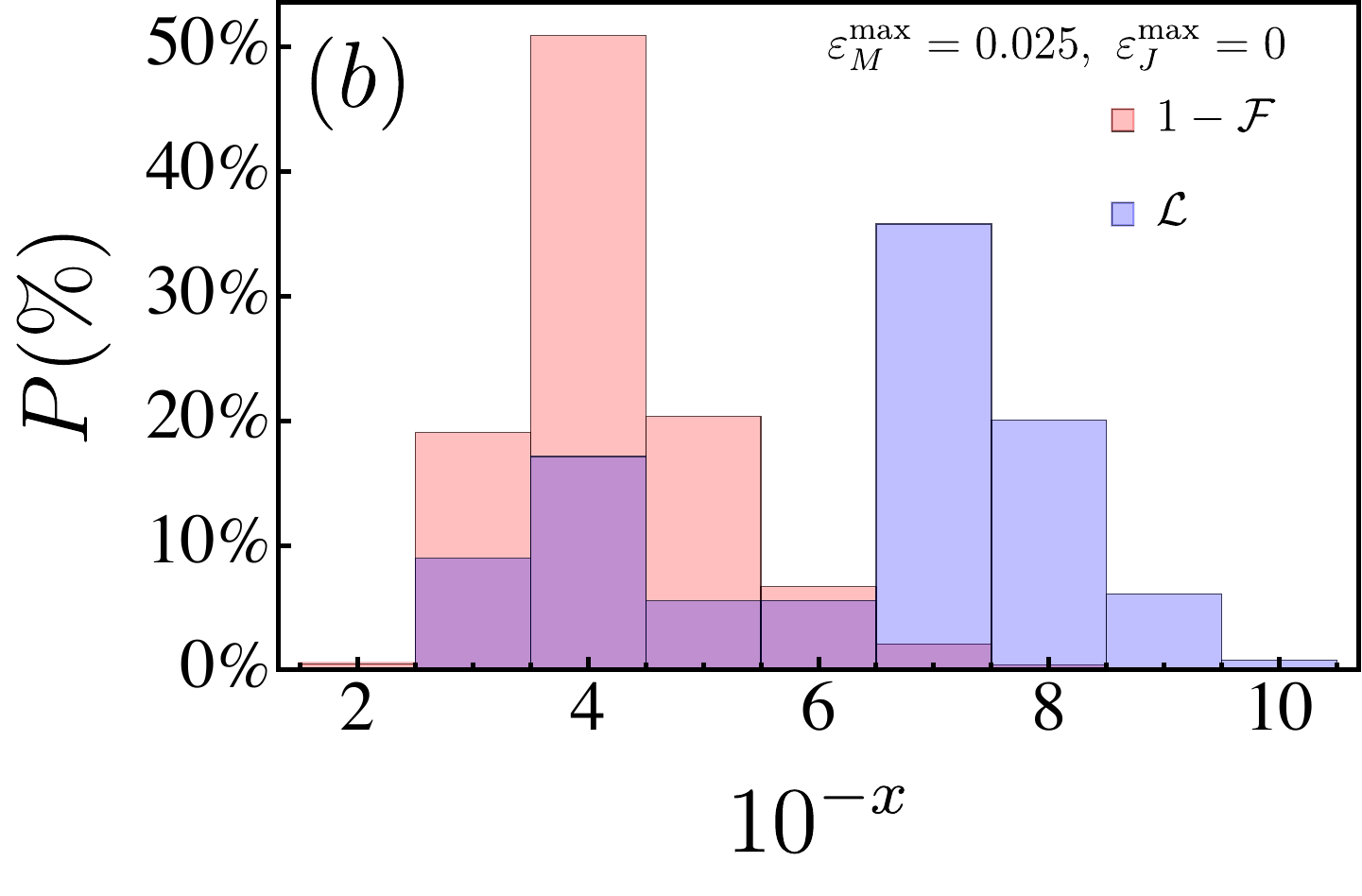}
	\includegraphics[width=0.48\linewidth]{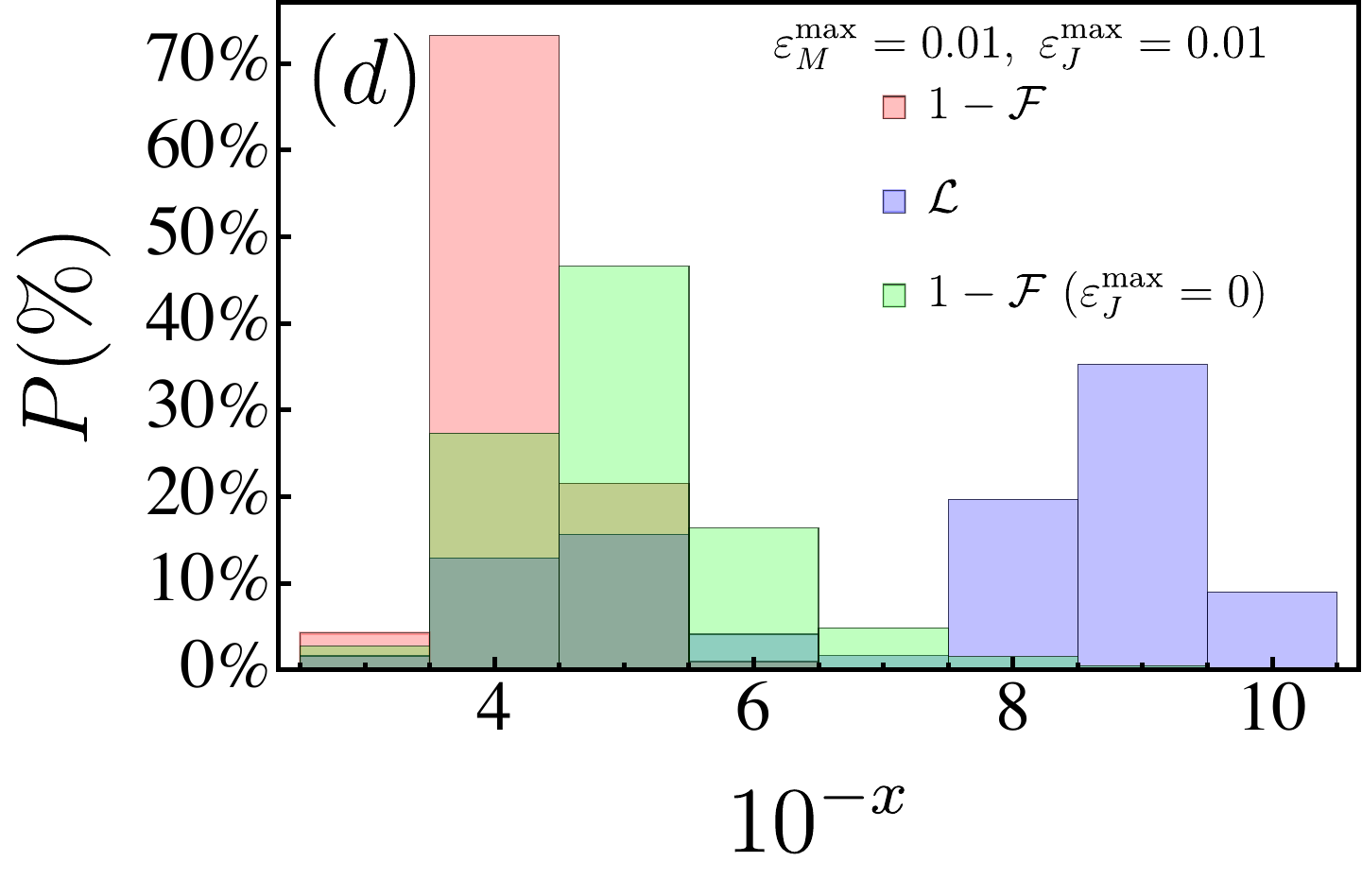}
	\caption{Histogram data of simulations of the proposed two qubit parity measurement based entangling gate with measurement fidelity errors, treating all measurement steps are measurements. Each data set is derived from $10^4$ iterations of the simulation and given values of maximum measurement fidelity $\varepsilon^{\text{max}}_M$ and exchange pulse $\varepsilon^{\text{max}}_J$ errors. (a) Order of magnitude distribution infidelity (red) and leakage (blue) of simulated errors $\varepsilon^{\text{max}}_M=1\%$ and $\varepsilon^{\text{max}}_J=0\%$. (b) Order of magnitude distribution of infidelity and leakage of simulated errors $\varepsilon^{\text{max}}_M=2.5\%$ and $\varepsilon^{\text{max}}_J=0\%$. (c) Order of magnitude distribution of infidelity and leakage of simulated errors $\varepsilon^{\text{max}}_M=5\%$ and $\varepsilon^{\text{max}}_J=0\%$. (d) Order of magnitude distribution of infidelity and leakage of simulated errors $\varepsilon^{\text{max}}_M=1\%$ and $\varepsilon^{\text{max}}_J=1\%$, including the infidelity of the simulation with $\varepsilon^{\text{max}}_J=0\%$ (green) for comparison.} 
	\label{fig:SimulationHistogramsLeak}
\end{figure}

The proposed two qubit gate given in Fig.~\ref{fig:two_qubit_Asym}(a) can been simulated to show expected average number of measurements, overall gate fidelity and possible leakage, accounting for infidelity in exchange gate and spin-parity measurements. Measurements in the simulation are projective, and treated as having random outcomes with the correct weighting. Initially, only exchange gate errors were considered, the results of these simulations are given in Fig.~\ref{fig:SimulationHistograms}. The simulation is run with a number of variants to the protocol and input states. Firstly, a simulation of the simplest experimental implementation is given in Fig.~\ref{fig:SimulationHistograms}(a), starting from an initial input state of $\ket{\psi_1 \psi_2}=\ket{SS}$ and ending in the state $\ket{\Psi^-}=(\ket{SS}-\ket{T_0 T_0})/\sqrt{2}$ after the implemented unitary. This possible experiment is discussed in greated detail in App.~\ref{sec:Experiment}. Here the distribution of the number of measurements needed when the final step to disentangle the ancilla is implemented as a measurement, is compared to when the disentangling is done by initialisation, so discussed in Sec.~\ref{sec:Asymmetric}. As expected, this shifts the distribution favorably toward fewer measurements and therefore a shorter overall gate time. Then the overall gate fidelity is probed by assuming some artificial small deflection of each exchange gate by a randomly selected $\varepsilon_J\in [-\varepsilon^{\text{max}}_J,\varepsilon^{\text{max}}_J]$. In Fig.~\ref{fig:SimulationHistograms}(b) the fidelity of the parity based gate with the final measurement and final re-initialisation are compared, along with a comparative fidelity distribution of a single phase gate of a random rotation angle acting on a random input state under an identically selected random error. Here, interestingly, the measurement based two qubit gates employing three or more error-prone exchange gates demonstrate a favorable fidelity distribution than a single phase gate. This is due to the gates reliance on interleaved projective measurement steps mitigating the accumulation of errors from the noisy exchange pulses. Lastly, the measurement based protocol is simulated with random input states in Fig.~\ref{fig:SimulationHistograms}(c) and (d) for a fuller characterisation of the average number of measurements and fidelity of the gate protocol.

From Fig.~\ref{fig:SimulationHistograms}(a) and (c), it is shown that the average number of measurements and exchange gates is equivalently between $3-5$ per execution of the proposed gate. Assuming parity readout times between $\unit[1-10]{\mu s}$\cite{seedhouse2020parity,philips2022universal}, entangling two qubit gate times of $\unit[5-50]{\mu s}$ can be achieved. While this is long compared to exchange gate times, $\sim\unit[10s-100s]{ns}$ for two qubit gates\cite{nichol2017high}, this is still shorter than the $T_1 \gg \unit[100]{\mu s}$\cite{philips2022universal} for single dot spin-$1/2$ qubits. While this may not seem like obvious point of comparison, as the qubits manipulated here are encoded here $S-T_0$ qubits, the interstitial measurements of the dots serve to effectively reset the spins in the processor such that a straightforward comparison of the gate time to the decoherence times of $S-T_0$ qubits is not a complete measure of quality. Instead, as the two outer dots of the $6$ dot device needed to perform such a gate are never measured, it is the decoherence times of these single spin-$1/2$ dots that limit the lifetime of a processor employing this control scheme. Thus the comparison of the proposed measurement based gate time with the $T_1 \gg \unit[100]{\mu s}$ for single spin-$1/2$ qubits is apt. 

There is no leakage induced in the proposed gate from simulated errors in the exchange pulses. To investigate possible leakage errors in the process, imperfect readout fidelities are included. These are modeled the same as errors in exchange pulses with some random small error $\varepsilon_M\in [-\varepsilon^{\text{max}}_M,\varepsilon^{\text{max}}_M]$ applied to each measurement, projecting onto the opposite outcome as is read out. The results of these simulations are given in Fig.~\ref{fig:SimulationHistogramsLeak}, assuming all measurement steps are treated as measurements (no singlet re-initialisation). In all the simulated values of $\varepsilon^{\text{max}}_M$, there are two clear peaks in the distribution of leakage. The peaks lower in the distribution of leakage ($10^{-8}-10^{-10}$) probability correspond on average to iterations of the simulated experiment where no repeat measurements are needed, whilst the second higher peak ($10^{-3}-10^{-6}$) correspond to iterations of the simulation where measurements repetitions are needed. Decreasing the readout fidelity (increasing $\varepsilon^{\text{max}}_M$) pushes these two peaks closer together. In Fig.~\ref{fig:SimulationHistogramsLeak}(d) errors in both measurements and the exchange pulses are compared to when perfect exchange gates are assumed with imperfect measurements. Here it is clear that the overall infidelity of the final state is perturbed most by the addition of the exchange pulse error from the large jump in the peak of the distribution of the infidelity. Overall the effect of leakage in the measurement based unitary is small, and is totally dependent on the fidelity of measurements used. Provided reported parity measurement fidelity of $99.98\%$\cite{philips2022universal}, the effect of leakage is effectively negligible.

\section{Discussion} \label{sec:Discussion}

In this work, measurements of joint-Pauli observables were discussed as a method of MBQC, demonstrating a straightforward recipe for designing single- and two- qubit gates. Combined with only one ancilla qubit, initialised to an eigenstate of a chosen Pauli observable, and a disentangling single qubit measurement, a unitary rotation on the data qubits are guaranteed by considered choice of measurements. Specifically, so long as each measurement steps' observable does not commute with that of the previous step, including the ancilla initialisation, the rotation will be unitary. Then, a physical parity observable in the form of spin-parity between two adjacent QDs was considered as a means to control DQD $\ket{S}-\ket{T_0}$ encoded qubits. Provided single-qubit phase gates by the natural spin-spin exchange interaction, a spin-parity measurement between spins from two neighbouring qubits is equivalent to an $XX$ parity measurement that can be tuned to a $YY$ parity measurement. The protocol is completely leakage-free with low overheads in terms of physical qubits, measurements and exchange gates. Then, a form of asymmetric spin-parity measurement shown in recent experiments was considered. Here, due to the projection of anti-symmetric spin states onto the maximally entangled $\ket{S_{L/R}}$ state, probabilistic coupling to leaked spin state is to be expected, but can be accounted for by the use of repeated measurements. Lastly, a two qubit parity-MBQC entangling gate is simulated with errors in the applied exchange pulses and measurements. The simulations show that the overall gate fidelity is robust against exchange infidelity, due to the projective measurements, and that leakage from measurement infidelity is effectively negligible.

Parity measurements of the form discussed also form a complete tool-set for quantum error correcting (QEC) codes. These include proposals for robust quantum memories via measurement-based QEC colour codes, as well as dynamically generated codes without a fixed logical basis. Our proposal for parity-MBQC with encoded spin qubits is compatible with such methods of QEC, and so is arguably scalable. Finally, it is believed that such gates are viable on current experimental spin qubit processors which offer spin-parity measurements and high fidelity control of the exchange interaction\cite{philips2022universal}. 

\section{Acknowledgments} \label{ref:Acknowledgements}
We acknowledge helpful discussions with U. G\"ung\"ord\"u and S. Hoffman.

\appendix

\section{Experimental Implementation}

\label{sec:Experiment}

\begin{figure}
	[t]
	\includegraphics[width=0.9\linewidth]{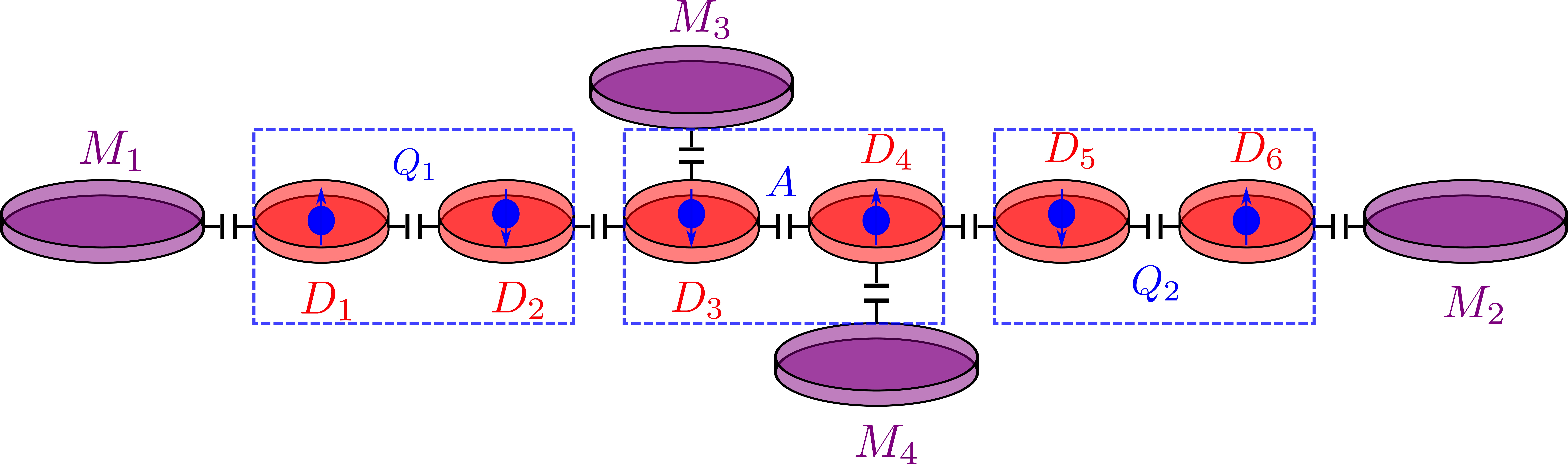}
	\caption{Example of a six dot device with which the proposed MBQC experiment could ideally be performed. The data dots in red, each charged with an electron in blue, are labeled $D_i$, and are grouped in pairs by the blue boxes labeling if they DQD pairs encode one of the two data qubits $Q_i$ or the ancilla $A$. The purple dots represent sensing (measurement) dots labeled $M_i$. In the case of a non ideal device that could still perform the proposed experiment (Ref.\cite{philips2022universal}), $M_3$ and $M_4$ and do not exist, and so layers of additional exchange $SWAP$ gates are needed to achieve the same protocol with just $M_1$ and $M_2$} 
	\label{fig:6DotDev}
\end{figure}

Here, details on how to perform a two qubit entangling gate by parity measurements on a linear six quantum dot device like the one in Ref.\cite{philips2022universal} is given. Ideally, in a device tailored as a small MBQC processor, sensing dots acting as quantum point contacts (QPCs) to read out changes in occupation of adjacent qubit dots would be positioned such that all the measurements needed to perform the desired gate set would be done so directly. In the case of a linear six dot device, encoding three DQD qubits whereby the outer two are data qubits and the middle is the ancilla, a minimum of four sensing dots is needed to directly access all necessary measurements. These include two at each end of the array to read out the occupation of the first and last dot, as well as two in parallel to the array above/below each one of the middle two dots to measure the occupation of both of the ancilla qubit dots. The outer sensing dots are used only to initialise and perform the final state readout on the two data qubits. This is done by Pauli spin-blockade style readout\cite{petta2005coherent}, tilting the detuning of the potentials toward the outer two dots to detect possible occupation change (a spin singlet state). The middle two sensing dots are assumed to be used for both initialisation and readout of the ancilla DQD qubit state as well as parity measurements between each dot of the ancilla qubit, and their neighboring data dot. Choice of measuring either the qubit basis (singlet-triplet) or spin-parity, in this case the perviously experimentally achieved asymmetric spin-parity measurement, depends on the choice of wait time at the measurement detuning\cite{seedhouse2020parity}. Such an ideal device is shown in Fig.~\ref{fig:6DotDev}.

The experiment described here is arguably the simplest proof-of-principal experiment of the two-qubit measurement based unitary described in Fig.~\ref{fig:two_qubit_Asym}. This version of the protocol takes the input states $\ket{\psi_1 \psi_2}=\ket{S S}=\ket{00}$ and performs a unitary rotation such that the output state is a maximally entangled Bell state $\ket{\Psi^-}=(\ket{00}-\ket{11})/\sqrt{2}$. Such an experiment simplifies the initialisation step of the protocol, and allows for some of the exchange pulses from the full arbitrary input implementation of the gate given in Fig.~\ref{fig:two_qubit_Asym}(a) to be omitted. However, if instead the two-qubit entangling gate is to be performed on arbitrary input states, these may be initialised in such a device by a combination of either starting with an initialised $\ket{S}$ or $\ket{+}$ depending on the how the $(0,2)\rightarrow(1,1)$ tunneling is handled\cite{philips2022universal}, exchange pulses, and applications of the measurement based single qubit gates shown in Fig.~\ref{fig:single_qubit_sym}.

The proof of concept experiment is as follows. Assuming a linear six dot array $D_1, D_2, \dots D_6$ with an ideal measurement setup of four sensing dots $M_1, M_2, M_3$ and $M_4$, an example two qubit entangling gate by measurement would be achieved as follows. Firstly, the three DQD encoded qubit states are prepared, in the case of the two data qubits, $Q_1 \rightarrow \{D_1, D_2\}$ and $Q_2 \rightarrow \{D_5, D_6\}$, into two identical spin singlet states by tunneling from a $(0,2)\rightarrow(1,1)$ charge configuration, as detected by the two sensing dots $M_1$ and $M_2$ coupled to qubit dots $D_1$ and $D_6$ respectively. This initialises a $\ket{\psi_{\text{init}}}=\ket{00}$ in the two qubit logical space. Equally, the ancilla DQD qubit $A \rightarrow \{D_3, D_4\}$ is prepared into a spin singlet state by monitoring the charge transition from either the intermediary sensing dots $M_3$ or $M_4$ coupled to $D_3$ or $D_4$ respectively. Once all the qubits are prepared then a $\pi/2$ pulse is applied between $D_1$ and $D_2$ and the first entangling measurement can be performed. In this case, the spin-parity between $Q_1$ and $A$ is checked. This is done by tilting the detuning potentials between $D_2$ and $D_3$ for the appropriate period of time such the sensing dot $M_3$ coupled to $D_3$ detects the desired parity. If the parity is detected to be anti-aligned ($\ket{S}$) then the measurement must be repeated after a $\pi$ exchange pulse is applied to $Q_1$. This is repeated until the spin aligned ($\ket{T_\pm}$) state is detected. Next, if the initial measurement of the parity of $Q_1$ and $A$ was spin aligned ($\ket{T_\pm}$) that a $\pi/2$ pulse is applied to $A$, else if any repeat measurements were needed, a $-\pi/2$ or $3\pi/2$ pulse is applied instead. Then the parity of dots $D_4$ and $D_5$ must be measured in the same manner as that of $D_2$ and $D_3$ with the sensing dot $M_4$ coupled to $D_4$. If the outcome of this second parity measurement matches that of the initial outcome of the first parity measurement, then a $\pi/2$ exchange pulse must be applied to $Q_1$, else if the outcomes are not equivalent a $-\pi/2$ or $3\pi/2$ exchange pulse is needed. Finally, the disentangling measurement can be done with either the sensor dots $M_3$ or $M_4$ coupled to $D_3$ of $D_4$ respectively, and must also probe the parity of the neighboring dots. Here as well, the application of up to one repeat measurements may be needed (see Tab.~\ref{tab:RepeatTab}). If the anti-aligned ($\ket{S}$) state is detected, the sequence is complete and no repetition is needed. Else if the outcome of the disentangling measurement is the spin anti-aligned ($\ket{T_\pm}$) state, then the measurement must be repeated by applying a $\pi$ exchange pulse between dots $D_4$ and $D_5$, as shown in Fig.~\ref{fig:two_qubit_Asym}(c), then measuring again. This should result in measuring the spin anti-aligned ($\ket{S}$) state, after which the sequence is complete. This measurement based gate sequence will leave the two qubits $Q_1$ and $Q_2$ in the maximally entangled Bell state $\ket{\Psi^-}=(\ket{SS}-\ket{T_0 T_0})/\sqrt{2}$, up to a known Pauli correction (see Fig.~\ref{fig:two_qubit_sym}(c)). The outcome can be verified by state tomography\cite{leon2021bell}, which can account for the Pauli correction. As previously discussed in Sec.\ref{sec:Asymmetric}, if the final measurement of the ancilla qubit is replaced with a initialisation by decay to the charge $(0,2)/(2,0)$ state, then the need for repetitions of the final measurement is eliminated, improving the average gate time of the protocol. Here however, nothing is changed by initialising instead of measuring the second parity measurement (between $D_4$ and $D_5$), as the final state is equivalent up to the Pauli corrections determined by this outcome. 

Finally, if, as in the case of Ref.\cite{philips2022universal}, the six dot device is purely linear, with no parallel sensing dots $M_3$ or $M_4$ coupled to $D_3$ and $D_4$ respectively, only those coupled to dots $D_1$ and $D_6$, then the same experiment can be achieved, only with sequences of $SWAP$ gates such the correct information is measured by the sensing dots. These swap gates are given by $\pi$ exchange pulses between neighboring dots. In the case of the first entangling measurement between qubit $Q_1$ and the ancilla $A$, a $SWAP$ gate sequence exchanging $D_1\longleftrightarrow D_3$ is needed such that the parity can be measured with the sensing dot coupled to $D_1$. Equally the second entangling measurement between qubit $Q_2$ and ancilla $A$ must employ a $SWAP$ sequence $D_4\longleftrightarrow D_6$ such that the sensing dot coupled to $D_6$ can probe the necessary parity. Then finally, a $SWAP$ sequence $D_1\longleftrightarrow D_3$, $D_3\longleftrightarrow D_5$ and $D_3\longleftrightarrow D_4$ is needed to effectively $SWAP$ qubit $Q_2$ with the ancilla $A$. Now the final disentangling measurement of the ancilla can be achieved with the sensing dot coupled to $D_6$.

\end{document}